\documentclass[prl,twocolumn,superscriptaddress,nobibnotes,a4paper]{revtex4-2}

\usepackage[pdftex]{graphicx}
\usepackage[pdftex]{epsfig}
\usepackage{amsmath}
\usepackage{amssymb}
\usepackage{amsfonts}
\usepackage{color}
\usepackage{wrapfig}
\usepackage{eucal}
\usepackage{hhline}
\usepackage{threeparttable}
\usepackage{supertabular}
\usepackage{multirow}
\usepackage{tabularx}
\usepackage{siunitx}
\usepackage{upgreek}
\usepackage{float}

\newcommand{\beven}{\beta_{\rm s}}
\newcommand{\bodd}{\beta_{\rm a}}
\newcommand{\NMS}{\Delta \nu_{\rm NMS}}
\newcommand{\FSR}{\Delta \nu_{\rm FRS}}
\newcommand{\coupling}{\beta}
\newcommand{\Ubs}{U_{\rm BS}}
\newcommand{\taurt}{\tau_{\rm rt}}

\usepackage[export]{adjustbox}

\usepackage{soul}

\usepackage[version=4]{mhchem} 
\usepackage{array} 
\usepackage[colorlinks=true, allcolors=blue]{hyperref} 

\setstcolor{red}

\begin{document}

\title{A single-phonon directional coupler}\thanks{This work was published in \href{https://doi.org/10.1364/OPTICAQ.569727}{Optica Quantum \textbf{3}, 445--451 (2025).}}

\author{Amirparsa Zivari}\thanks{These authors contributed equally to this work.}
\affiliation{Kavli Institute of Nanoscience, Department of Quantum Nanoscience, Delft University of Technology, 2628CJ Delft, The Netherlands}
\author{Niccol\`{o} Fiaschi}\thanks{These authors contributed equally to this work.}
\affiliation{Kavli Institute of Nanoscience, Department of Quantum Nanoscience, Delft University of Technology, 2628CJ Delft, The Netherlands}
\author{Lorenzo Scarpelli}\thanks{These authors contributed equally to this work.}
\affiliation{Kavli Institute of Nanoscience, Department of Quantum Nanoscience, Delft University of Technology, 2628CJ Delft, The Netherlands}
\author{Menno Jansen}
\affiliation{Center for Nanophotonics, AMOLF, Science Park 104, 1098XG Amsterdam, The Netherlands}
\affiliation{Department of Applied Physics and Science Education and Eindhoven Hendrik Casimir Institute, Eindhoven University of Technology, P.O.\ Box 513, 5600MB Eindhoven, The Netherlands}
\author{Roel Burgwal}
\affiliation{Center for Nanophotonics, AMOLF, Science Park 104, 1098XG Amsterdam, The Netherlands}
\affiliation{Department of Applied Physics and Science Education and Eindhoven Hendrik Casimir Institute, Eindhoven University of Technology, P.O.\ Box 513, 5600MB Eindhoven, The Netherlands}
\author{Ewold Verhagen}
\affiliation{Center for Nanophotonics, AMOLF, Science Park 104, 1098XG Amsterdam, The Netherlands}
\affiliation{Department of Applied Physics and Science Education and Eindhoven Hendrik Casimir Institute, Eindhoven University of Technology, P.O.\ Box 513, 5600MB Eindhoven, The Netherlands}
\author{Simon Gr\"oblacher}
\email{s.groeblacher@tudelft.nl}
\affiliation{Kavli Institute of Nanoscience, Department of Quantum Nanoscience, Delft University of Technology, 2628CJ Delft, The Netherlands}


\begin{abstract}
Integrated photonics has revolutionized fields such as telecommunications, quantum optics, and metrology by enabling compact, scalable circuits through highly confined optical modes. Within the field of quantum acoustics, phonons have emerged as a compelling alternative, offering advantages such as lower energy, smaller mode volume, and low propagation speeds, which make them ideal for interfacing diverse quantum systems. Developing integrated phononic circuits is thus essential for unlocking the full potential of quantum acoustics and advancing technologies such as quantum computing and hybrid systems. In this work, we demonstrate the first 4-port directional coupler for quantum mechanical excitations – a key building block for phononic circuits. By tuning the coupling region length, we achieve phononic beam splitters with controllable splitting ratios. We validate quantum-level performance by sending a single–phonon Fock state through the device. This work represents a foundational advance toward scalable, integrated phononic platforms for both classical and quantum applications.
\end{abstract}
	
\maketitle
	
\section*{Introduction}
Quantum technologies have seen rapid progress in the past few years, with great promise for testing fundamental science, as well as for commercial applications. First attempts to achieve quantum computational advantage~\cite{Arute2019,ZhongS2020} over classical processors and long-distance quantum teleportation using a satellite~\cite{Ren2017} have highlighted the fundamental need to create heterogeneous quantum systems~\cite{Wallquist2009} in order to realize advanced quantum technologies~\cite{Kurizki2015}, such as quantum networks~\cite{Kimble2008}. Phonons -- quantized mechanical vibrations -- are regarded as a critical resource to connect different quantum devices~\cite{Delsing2019,Barzanjeh2022}, with applications in, for example, quantum transduction~\cite{Mirhosseini2020,Jiang2020, Weaver2023} and sensing~\cite{Fogliano2021}. Accordingly, routing and manipulating single mechanical vibrations on a chip is crucial to transfer quantum information between different quantum systems and unlock the potential of hybrid quantum systems.

\begin{figure*}[ht!]
	\includegraphics[width = 1.0 \linewidth]{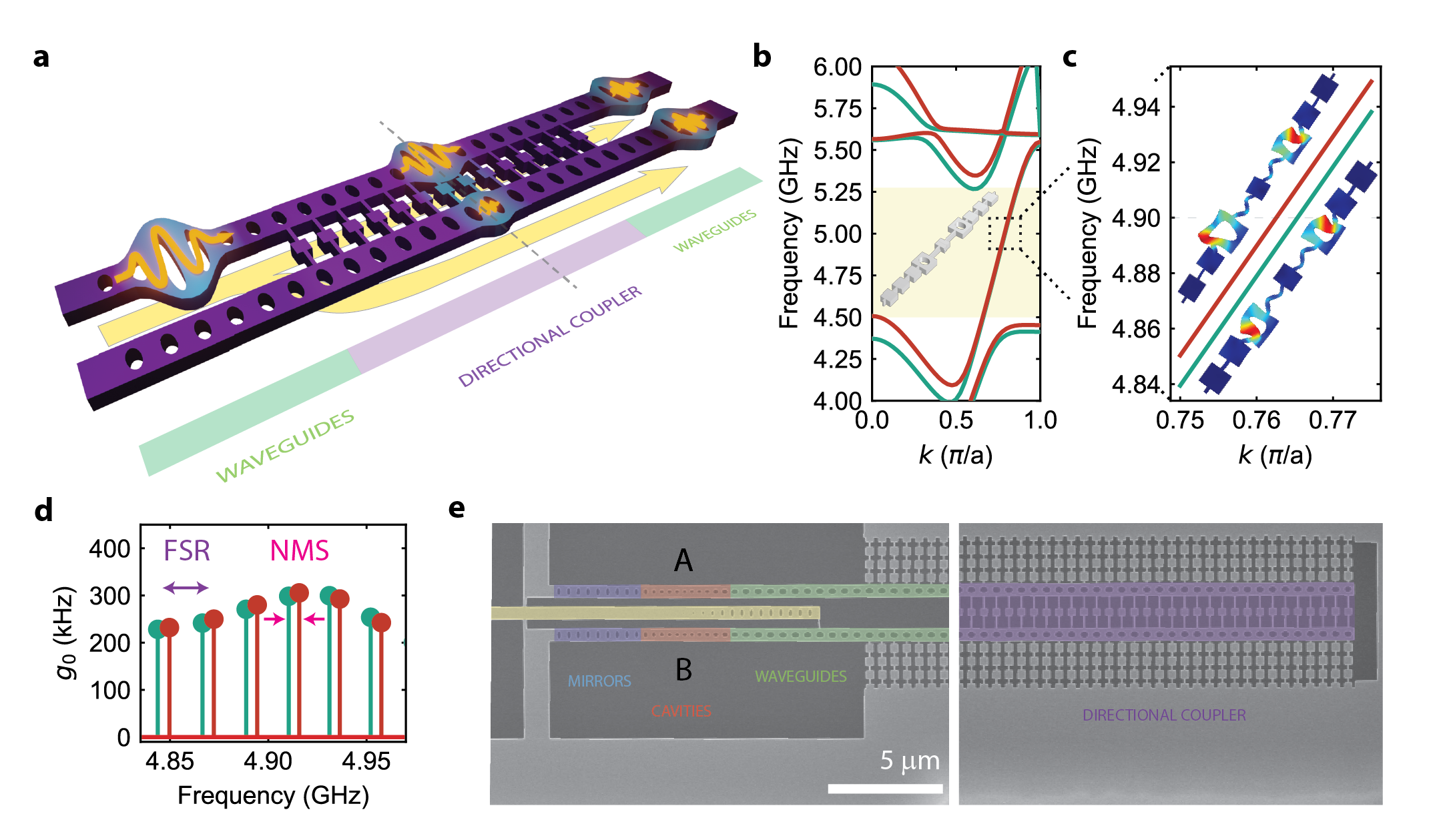}
	\caption{a) Sketch of the phononic directional coupler device, which includes two uncoupled phononic waveguides (green) coupled together in the central region (purple). The orange pulses represent classical wavepackets. A wavepackets generated in one of the waveguide splits into two identical wavepackets after transmission through the coupler with a 50:50 spitting ratio. b) Band structure simulation of the in-plane breathing mode of the unit cell structure (inset) of the directional coupler, showing normal mode splitting of the symmetric (green line) and anti-symmetric (red line) supermodes due to the coupling. c) The splitting between the two modes is clearly visible in the zoom-in around \SI{4.9}{GHz}. The oscillating anti-symmetric and symmetric supermodes are shown on the top and bottom, respectively. d) Simulated optomechanical single photon coupling rate of the full structure. The mechanical mode of the optomechanical cavity is hybridized by a series of Fabry-P\'erot modes of the waveguide and each mode is split into a doublet of symmetric (green) and anti-symmetric normal modes (more details are given in SI~\ref{SI:sim}. e) Scanning electron microscope (SEM) images of the fabricated device with false color highlighting of the different parts -- the phononic and photonic mirrors (blue), the optomechanical cavities (red), the phononic waveguide (green), the phononic directional coupler (purple). For a 50:50 splitting, the coupler length is about \SI{26}{\mu m}, and the total device length is about \SI{210}{\mu m}. Due to space constraints, only a short section of the waveguides is shown.}
	\label{Fig:1}
\end{figure*}

However, to date, for quantum information purposes based on traveling phonons, two main platforms have been developed:\ surface acoustic waves (SAWs)~\cite{Bienfait2019,Bienfait2020,McNeil2011} and highly confined one-dimensional phononic waveguides based on phononic crystals~\cite{Zivari2022}. In particular, SAWs have been used to couple two qubits to one another over a distance of around 2\,mm~\cite{Bienfait2019}. The propagation distance of these acoustic waves is however limited by the relatively short phononic lifetime $T_{1}$ in the range of tens of microseconds. At the same time, due to the two-dimensional, inherently open nature of SAWs, devices are characterized by a relatively large footprint, making scaling to more complex circuits challenging. Nevertheless, SAW platforms are suitable candidates for coupling single traveling phonons to superconducting qubits, which are deterministic sources and detectors for single phonons, and have been used in  various outstanding studies in quantum acoustics.
Highly confined traveling GHz phonons in phononic crystal waveguides on the other hand are characterized by milliseconds-long lifetimes~\cite{Zivari2022} and feature a relatively small footprint compared to SAWs, making them an ideal platform for on-chip quantum applications~\cite{zivari2022chip} and for integrated phononic circuits in general. In these structures, an opto-mechanical resonator is typically used for the generation, manipulation, and detection of highly confined phonons:\ from realizing single Fock states of phonons~\cite{Riedinger2016,Hong2017}, remote phonon-phonon quantum entanglement~\cite{Riedinger2018}, and optomechanical Bell tests~\cite{Marinkovic2018} to quantum teleportation~\cite{Fiaschi2021}. More recently, single-phonon waveguides were successfully connected to such opto-mechanical resonators to route single phonon wavepackets on a chip~\cite{zivari2022chip}.

In photonics, beam splitters form one of the fundamental building blocks for a myriad of applications -- they provide a classical platform to investigate the wave nature of light, are used as combiners and power distributors, are critical for quantum optics experiments, and are an essential resource for linear optical quantum computing and processing~\cite{knill2001scheme}, to name a few examples. Importantly, the ability to miniaturize photonic components has been paramount for realizing novel technologies, as it enables photonic integrated circuits with thousands of components while still preserving a millimeter-scale footprint~\cite{wang2020integrated,liu2023aluminum,wang2023photonic}. Some classical demonstration of beam-splitters for GHz phonons have been realized over the past years, showing the growing interest in this topic~\cite{Mayor2021,Zhang2022,Feng2023}. Additionally, significant progress has recently culminated in a remarkable proof-of-principle demonstration of a phononic beam splitter for SAWs~\cite{qiao2023splitting} in quantum regime. In close analogy to photonics, here we experimentally realize the phononic equivalent of this crucial component -- an integrated phononic beam splitter -- one of the most critical elements to perform advanced on-chip manipulation of mechanical excitations, in both classical and quantum regime. Drawing inspiration from photonics, we design our beam splitter using a 4-port directional coupler architecture, where two identical single-mode waveguides are coupled together in an interaction region, creating symmetric and anti-symmetric supermodes which are delocalized in both waveguides. The splitting ratio can be adjusted by changing the length of this interaction region. We demonstrate the beam splitter behavior for both coherent and single-phonon Fock states. With a footprint of only about $200\times5~\mu$m, our device is easily scalable and represents a critical step towards the realization of integrated phononic circuits~\cite{Fang2016,Patel2018,MadiotPRL2023}.

\section*{Device Design and Characterization}
A conceptual sketch of our device is shown in Fig.~\ref{Fig:1}a. The device is formed by two single-phonon sources (the optomechanical cavities), two single-mode waveguides, a directional coupler, and two single-phonon detectors. To realize the phononic beam splitter, we design an integrated directional coupler using a phononic crystal architecture. The corresponding unit cell of length $a$ is shown in the inset of Fig.~\ref{Fig:1}b. It consists of two single-mode waveguides~\cite{Zivari2022}, connected together via a phononic bridge. The phononic bridges have a full bandgap at the frequencies of interest. Therefore the coupling is evanescent, and is just a weak perturbation to the uncoupled modes. This interaction region allows mechanical energy to be exchanged between the waveguides, resembling the evanescent coupling typically used in photonic directional couplers. In contrast to a conventional directional coupler, the two waveguides host both the two incoming, as well as the two outgoing ports, as detailed below. To verify the analogy to a photonic directional coupler, we perform finite element simulations of the unit cell using COMSOL and calculate the corresponding band structure. The results are shown in Fig.~\ref{Fig:1}b, for the breathing mode in the in-plane direction. Within a frequency range of $4.5-5.3$~GHz, the band structure is dominated by the symmetric and anti-symmetric supermodes, which are the even and odd linear combinations of the uncoupled modes, respectively. This range also defines the operational bandwidth of the current design. A zoom-in around \SI{4.9}{GHz} shown in Fig.~\ref{Fig:1}c reveals that the two supermodes have different propagation constants $\beta_{\rm s,a}$, which relate, according to conventional coupled mode theory~\cite{huang1994coupled}, to the coupling coefficient $\beta$ per unit length as $\beta = (\beven - \bodd)/2$, with the subscripts $i={\rm \{s,a\}}$ indicating the symmetric and anti-symmetric supermodes, respectively. As a result of the coupling, for each wave vector, the supermodes are frequency split by the normal mode splitting $\NMS$, as expected from two coupled degenerate harmonic oscillators. We use the extracted value of $\coupling$ to estimate the coupling length needed for different splitting ratios.

To reduce the experimental complexity, we engineer a device with a (phononic) mirror at the center of Fig.~\ref{Fig:1}a (represented by the dashed line, and realized by a free-standing end)~\cite{Zivari2022,zivari2022chip}. In this way the same two cavities can be used as transmitters and receivers. We note that, for temporally distinguishable wavepackets as is the case in our experiments, this is equivalent to having a physical 4-port system. The output arms of the directional coupler are connected to two uncoupled, single-mode, waveguides, each terminated with a (nominally identical) optomechanical cavity. The cavities are used to generate and detect single phonons via the optomechanical interaction. Through finite element simulation of the full structure we can calculate the optomechanical coupling rates for different modes, plotted in Fig.~\ref{Fig:1}d. The optomechanical response is dominated by a series of Fabry-P\'erot modes, equally spaced by $\FSR$, which arise from the hybridization of the mechanical cavity mode with the series of modes supported by the free-ending waveguide. Furthermore, each Fabry-P\'erot mode is split by $\NMS$ into a doublet due to the mechanical coupling between the two devices (see SI~\ref{SI:sim} for more details). Scanning electron microscope (SEM) pictures of one of the fabricated devices, made from a \SI{250}{nm} thick device layer of an SOI wafer~\cite{Zivari2022,zivari2022chip}, are shown in Fig.~\ref{Fig:1}e. For space reasons the waveguides are not fully shown and the total length of the device is about \SI{200}{\mu m}.

\begin{figure}[]
	\includegraphics[width = 1 \linewidth]{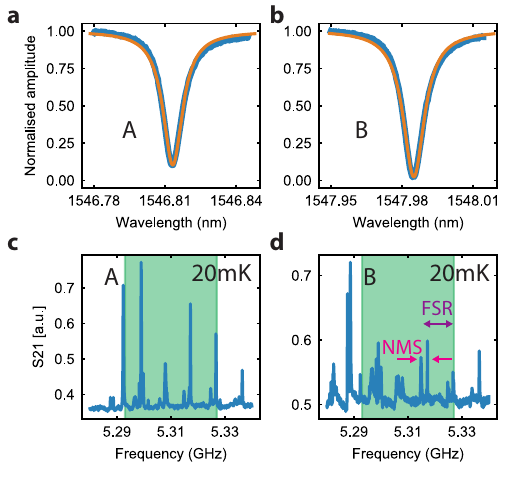}
	\caption{a,b) Optical response of device A (left) and B (right). The devices have optical resonances in the C-band at telecom wavelengths. c,d) Mechanical spectrum of device A (left) and B (right) measured using OMIT. We observe a series of Fabry-P\'erot modes spaced by $\FSR$ and split by the normal-mode splitting $\NMS$ due to the directional coupler. The green shaded area indicates the spectral region used to investigate the dynamics of phononic wavepackets (see text for more details).}
	\label{Fig:2}
\end{figure}

To minimize the thermal noise background, we cool our device to \SI{20}{mK} using a dilution refrigerator, initializing all mechanical modes of interest in their quantum ground states. We measure the optical characteristics of the two optomechanical cavities by scanning a continuous-wave laser through their resonances and measure the reflected signal on a photodiode. The reflection spectra are shown in Fig.~\ref{Fig:2}a and b (for cavity A and B, respectively). Fitting the line shape with a Lorentzian, we determine the optical cavity resonance at $\lambda \approx \SI{1546.81}{nm}$ ($\lambda \approx \SI{1547.98}{nm}$), and a full-width at half-maximum (FWHM) of $\kappa/2\pi \approx \SI{1.23}{GHz}$ ($\kappa/2\pi \approx \SI{1.34}{GHz}$), with an intrinsic loss rate of $\kappa_i/2\pi \approx \SI{430}{MHz}$ ($\kappa_i/2\pi \approx \SI{600}{MHz}$), for device A (B), respectively. The cavities are nominally identical, apart from fabrication imperfections. Since the difference between the optical resonances is much bigger than the linewidths and mechanical frequencies, we can address each cavity individually. This allows us to measure the mechanical spectrum of the structure using optomechanically induced transparency (OMIT)~\cite{Weis2010}. The results are plotted in Fig.~\ref{Fig:2}c and d (for cavity A and B, respectively). Both devices, within a region of $\sim \SI{30}{MHz}$ (highlighted in green shaded area) show a series of doublets with a normal mode splitting of $\NMS \approx \SI{2.5}{MHz}$, spaced almost evenly by $\FSR \approx \SI{10}{MHz}$. As will be shown later, this is a signature of mode hybridization through the directional coupler, and corresponds to a directional coupler with a splitting ratio of approximately 50:50. We further determine the equivalent single photon optomechanical coupling rate from the Stokes scattered photon rate by sending \SI{30}{ns} optical pulses to the devices. We estimate $g_{0,A} = \SI{380}{kHz}$ and $g_{0,B} = \SI{530}{kHz}$ for device A and B, respectively. We note that these are effective coupling rates, resulting from the coupling rate of all modes excited within the pulse length. Furthermore, the simulated coupling rates shown in Fig.~\ref{Fig:1}d are calculated for a shorter device length compared to the measured device (for faster computational run-times) and can therefore not be directly compared to the experimental values.

\section*{Coherent drive measurement}
In order to characterize the time-domain behavior of the beam splitter, we study the propagation of coherent phononic wavepackets. We detune the laser by \SI{6}{GHz} from the optical cavity we aim to excite. By using this detuning the Stokes (and anti-Stokes) scattering rate are strongly suppressed and can be neglected. We then modulate sidebands at the mechanical frequency $f_m = \SI{5.31}{GHz}$ onto the laser using an electro-optical modulator (EOM). The sidebands create a beating tone at the mechanical frequency, which coherently drives the mechanical motion through radiation pressure. The optical pulses are created using an acousto-optical modulator and are \SI{30}{ns} long, such that we excite modes only within the spectral region shown in the green shaded area in Fig.~\ref{Fig:2}c and d, for cavity A and B, respectively.
	
\begin{figure}[]
	\includegraphics[width = 1 \linewidth]{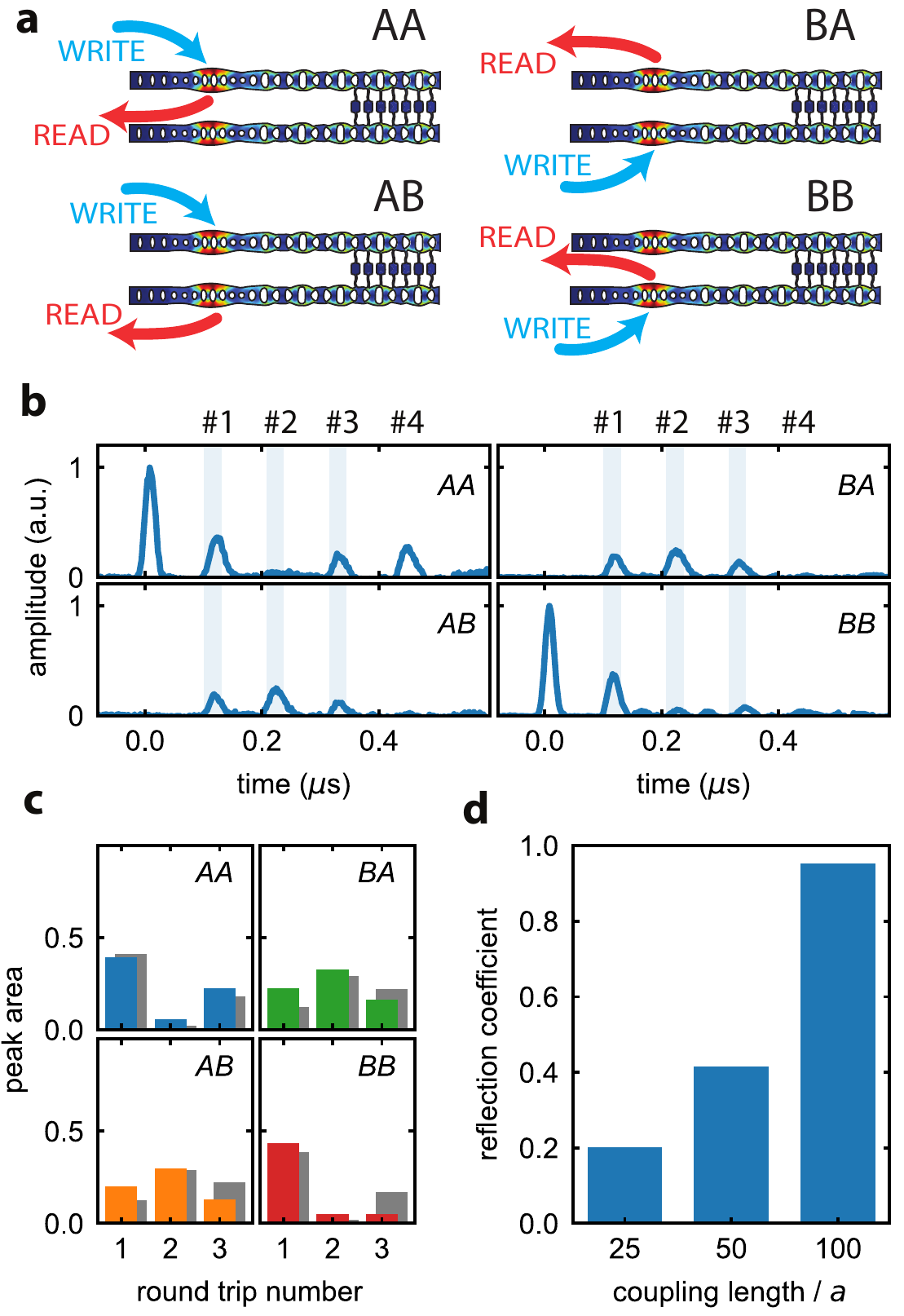}
	\caption{a) Different permutations of exciting and reading the two cavities. b) Normalized phononic population in different cavities with respect to the excited population at t = 0 (first letter:\ excited cavity, second letter:\ read-out cavity), using a coherent state phononic wavepacket. The number on the top corresponds to the round trip number of the phononic packet with the respective time range highlighted in blue. c) Blue, orange, green, red color bars:\ integrated counts of phononic population around the first, second and third round-trip time, normalized to the amplitude of the peak at zero delay in the excitation waveguide. Grey color bars:\ result of the global fit described in the main text, from which we extract a reflection coefficient $R=0.43$. d) The reflection coefficient of the phononic directional coupler as a function of the coupling length, measured on different devices.}
	\label{Fig:3}
\end{figure}

To measure the phonon occupancy inside the cavities A and B, we again use a continuous-wave laser, red-detuned by the mechanical frequency from the optical resonances. The results of the normalized phononic population, with respect to the excited population, in the cavities over time are shown in Fig.~\ref{Fig:3}b, for different combinations of exciting (first letter on the top right corner of each panel) and reading (second letter on the top right corner of each panel) cavities, as it is depicted in Fig.~\ref{Fig:3}a -- see SI~\ref{SI:coh_drive} for more details. This device has a coupler length of 50 unit cells, which corresponds to an approx.\ 50:50 beam splitter. As a result of the directional coupler, the excitation is then split between the two waveguides and reaches the cavities A and B after the first round trip, with a $\pi/2$ phase difference between the arms. Subsequently, at the second recombination on the coupler region (second round-trip), the excitation travels to the opposite port with respect to the one excited initially. This is due to the fact that the phase difference between our phononic packets in the two arms remains constant while they travel through the waveguides~\cite{zivari2022chip}. Due the slight mismatch of the two waveguide-cavity systems and dispersions in the coupler region caused from the fabrication disorders, the amount of energy populating the opposite cavities is slightly different. We note that despite a longer lifetime in these confined structures, the phononic population dims out after about 700~ns. This is mainly caused by dispersion and mismatch between the two waveguides rather than phononic dissipation. Nevertheless, this does not impose a fundamental limit for loss and can be improved by designing a phononic crystal more robust to disorders and by improving the overall fabrication precision~\cite{zivari2022chip, Zivari2022}.
	
To extract the splitting ratio of the directional coupler, we calculate the (normalized) area under the mechanical wavepacket around the first, second, and third round-trip time, corresponding to the light-blue shaded area in Fig.~\ref{Fig:3}b, for the indicated excitation and detection combinations (see SI~\ref{SI:coh_drive} for more details). The corresponding integrated counts are shown in blue, orange, green and red color bars in Fig.~\ref{Fig:3}c. To model these data, we use a transfer matrix approach, where the action of the directional coupler is described by a lossless beam splitter matrix $\Ubs$, written in terms of a reflection coefficient $r$ and a transmission coefficient $t = \sqrt{1-r^2}$. Phonon losses are negligible to first order since the phononic lifetime is much longer than the round-trip time (see SI~\ref{SI:lifetime}). The phonon population after $N$ passes through the directional coupler is then proportional to $A_{ij}\left(\Ubs\right)^N$, where the factor $A_{ij}$, with $i,j = \{{\rm A,B}\}$, takes into account losses and dispersion mismatch. A more accurate description of these terms would require modeling of the time dynamics with a multi-mode coupled-mode theory, which goes beyond the scope of this work. We then perform a global fitting procedure, where we impose $A_{\rm AB}=A_{\rm BA}$ to satisfy the reciprocity condition. The result of the fit is shown in grey color bars in Fig.~\ref{Fig:3}c, from which we extract a reflection coefficient $R=|r|^2=0.43$. We repeat the same measurements for two different lengths of the coupler region (see SI~\ref{Fig:coh_drive_SI}) and extract the corresponding reflection coefficient. The results are shown in Fig.~\ref{Fig:3}d, which clearly shows an increase of the reflection as the coupling length is increased. Importantly, these results show that our design allows to engineer phononic beam splitters with arbitrary splitting ratios. We note that we measure negligible reflections at the interface between the uncoupled waveguides and the coupler region (see SI~\ref{SI:coh_drive} for more details).

\section*{Single phonon splitting}
Due to the relatively low mode frequencies of phonons compared to optical photons in general, and heating induced by optical absorption in optomechanical systems in particular, thermal noise can play a major role in phononic systems, reducing the purity of quantum states, such as single-phonon Fock states. Therefore, going beyond classical demonstrations \cite{Feng2023} and testing the response of our directional coupler to quantized excitations with single phonon wavepackets is a crucial demonstration of its potential use for quantum applications. This becomes even more critical when considering additional phononic losses, for example via interaction with two-level systems, as well as traveling losses due to dispersion, as all these effects decrease the phononic population compared to the thermal occupancy, thus effectively augmenting decoherence. In this experiment, we excite single phonon wavepackets in different cavities and detect the reflected signals in both cavities. For a single phonon input state incident on a beam splitter with a 50:50 splitting ratio, there is an equal 50\% chance of receiving a single phonon on either of the output ports.

In order to create a single phonon, we use \SI{30}{ns} ``write" laser pulses, blue-detuned by the central mechanical frequency (\SI{5.31}{GHz}) from the optical resonances. This process can be described by a two-mode squeezed optomechanical interaction and detecting the Stokes scattered photon with a superconducting nanowire single photon detector (SNSPD), allows us to project the mechanical state of the optomechanical cavity onto a single phonon state~\cite{Riedinger2016}. This single phonon travels through the waveguide and either returns to the same cavity upon reflection, or goes to the opposite cavity after transmission in the coupling region. We then convert the mechanical state of each cavity to an optical photon, by sending \SI{30}{ns} ``read" laser pulses red-detuned by the same mechanical frequency from the optical cavities. This addresses a state swap optomechanical interaction resulting into anti-Stokes scattered photons. The time delay between the write and read pulses is set around the first round trip of the phononic wavepacket. We calculate the second order cross-correlation $g^{(2)}_{om}$ between the Stokes and anti-Stokes scattered photons for different combinations of exciting and reading cavities A and B, similarly to~\cite{Riedinger2016}.

\begin{figure}[]
	\includegraphics[width = 1 \linewidth]{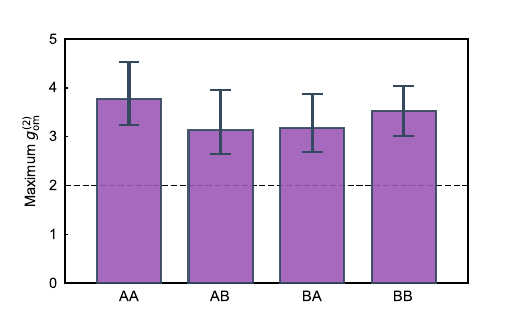}
	\caption{The second order cross-correlation $g^{(2)}_{om}$ between the Stokes and anti-Stokes scattered photons from the write and read pulses -- we measure $g^{(2)}_{om,AA} = 3.8^ {+0.8}_{-0.5}$, $g^{(2)}_{om,AB} = 3.1^ {+0.8}_{-0.5}$, $g^{(2)}_{om,BA} = 3.2^ {+0.7}_{-0.5}$ and $g^{(2)}_{om,BB} = 3.5^ {+0.5}_{-0.5}$ -- all values are above the classical threshold of 2 (dashed line) by more than two standard deviations. Error bars indicate one standard deviation.}
	\label{Fig:4}
\end{figure}

We use pulse energies of \SI{220}{fJ} for write (Stokes) and \SI{280}{fJ} for read (anti-Stokes) process, corresponding to a Stokes scattering probability of \SI{1}{\%} (\SI{1.6}{\%}) and an anti-Stokes scattering probability of \SI{1.2}{\%} (\SI{2}{\%}) for cavity A (B). At these low pulse energies, the thermal occupation of the cavities caused by heating of the lasers is negligible ($n_{th}<0.18$ for all measurements configuration, see SI~\ref{SI:nth}). We measure values of $g^{(2)}_{om,AA} = 3.8^ {+0.8}_{-0.5}$, $g^{(2)}_{om,AB} = 3.1^ {+0.8}_{-0.5}$, $g^{(2)}_{om,BA} = 3.2^ {+0.7}_{-0.5}$ and $g^{(2)}_{om,BB} = 3.5^ {+0.5}_{-0.5}$ (first index indicating the write cavity and second index the read cavity), which is plotted in Fig.~\ref{Fig:4}. For each configuration the value is more than two standard deviations above the classical threshold of 2, unambiguously showing the non-classical behavior of the single phonon states measured in the different cavities after the excitations passed through the coupler region and either return back to the same cavity or travel to the opposite one. We note that these values are consistent with the expected ones for the first round trip given the thermal occupation and the splitting ratio:\ $g^{(2)}_{om, exp} \approx (1+\alpha/n_{th})\approx 3.7$, with $\alpha = R$ ($T$) for $AA$, $BB$ ($AB$, $BA$)~\cite{Riedinger2016,Zivari2022}. For more details on the experimental setup we refer to SI~\ref{SI:gcc} and SI~\ref{SI:setup}. The equal $g^{(2)}_{om}$ values for different combinations of excitation and detection cavities, proves the equal chance of retrieving the phonon in different output ports (i.e.\ $R = T$), and therefore the 50:50 splitting of a single phonon state.

A natural next step would be to perform a Hong-Ou-Mandel interference experiment. However, due to the low clickrates, the corresponding integration time would currently be prohibitively large. Using different types of optomechanical cavities or a deterministic source of phonons could allow for this experiment~\cite{MacCabe2020,qiao2023splitting}.

\section*{Conclusion}
In conclusion, we have demonstrated an integrated directional coupler for GHz phonons -- a crucial component for phononic integrated circuits~\cite{Fu2019}. Our demonstration is based on the development of the first integrated circuit involving single-phonon sources, detectors and single-mode waveguides, and we use it to demonstrate a beam splitter for both classical and quantum mechanical states. Importantly, the circuit is fabricated with only one lithographic step. Due to the millisecond-long lifetime of phonons in these structures, as well as the ease of scalability, our device can be readily extended to build multi-mode interferometers of large dimensions, suitable for mechanical boson sampling and, more generally, for linear mechanical quantum computing~\cite{ZhongS2020,qiao2023splitting}. We would like to note that due to having a passive cooldown/reset protocol for our mechanical oscillators, longer lifetimes would lead to a decrease in the repetition rate of the pulse sequence. Therefore, in order to have reasonable data acquisition times in the milliseconds lifetime regime, the scattering probabilities and click rates must also be increased. This is only possible by having optomechanical devices with lower optical absorption to be able to increase the optical pump powers~\cite{Wallucks2020, Ren2019, Hong2017, Borselli2006, chen2024bandwidth}.

Through further development, piezo-electric materials can be included in our mechanical structure~\cite{Jiang2023,Weaver2023}, enabling electro-mechanical integrated devices such as, for example, phononic phase shifters~\cite{taylor2022reconfigurable}. With this powerful new tool at hand, a new paradigm of phononic devices can be realized, including mechanical Mach-Zender interferometers for sensing, switches for routing, and power multiplexers, to name a few, in addition to other more versatile mechanical counterparts of existing photonic technologies~\cite{Tian2021}.

Additionally, using piezo-electric resonators, strong interactions between highly-confined GHz phonons and superconducting qubits can be engineered~\cite{Mirhosseini2020}. Moreover, the small mode-volume of highly confined phonons achieved with our design, allows engineering interactions with nanometer-scale quantum system, such as quantum dots~\cite{Spinnler23} and color centers. Therefore, our directional coupler opens up exciting perspectives for hybrid quantum networks, enabling direct entanglement generation between distinct quantum systems without requiring entanglement swapping. We envision a platform with several quantum resources linked and combined together through mechanical quantum channels, taking advantage of the best performances of each individual technology. Our design, based on one-dimensional waveguides, resembles the one of a photonic integrated circuit. Together with the demonstration of non-classical behavior, our work paves the way towards expanding the new field of integrated quantum phononics.
\\

\section*{Acknowledgments}
We would like to thank the Kavli Nanolab Delft for nanofabrication assistance. This work is financially supported by the European Research Council (ERC CoG Q-ECHOS, 101001005) and is part of the research program of the Netherlands Organization for Scientific Research (NWO), supported by the NWO Frontiers of Nanoscience program, as well as through a Vrij Programma (680-92-18-04) grant.

\textbf{Conflict of interests:}\ The authors declare no competing interests.

\textbf{Author contributions:} A.Z., L.S., N.F.\ and S.G.\ devised and planned the experiment. A.Z., N.F, M.J., R.B.\ simulated and designed the device. A.Z.\ and L.S.\ fabricated the sample, A.Z., L.S.\ and N.F.\ built the setup and performed the measurements. A.Z., L.S., and S.G.\ analyzed the data and wrote the manuscript with input from all authors. E.V.\ and S.G.\ supervised the project.

\textbf{Data Availability:}\ Source data for the plots are available on \href{https://zenodo.org/records/17158678}{Zenodo}.

\setcounter{figure}{0}
\renewcommand{\thefigure}{S\arabic{figure}}
\setcounter{equation}{0}
\renewcommand{\theequation}{S\arabic{equation}}

\clearpage

\begin{center}
	\textsc{\Large Supplementary Information} 
\end{center}
\label{SI}

\section{Simulation}
\label{SI:sim}	
\begin{figure*}[ht!]
	\includegraphics[width = 1 \linewidth]{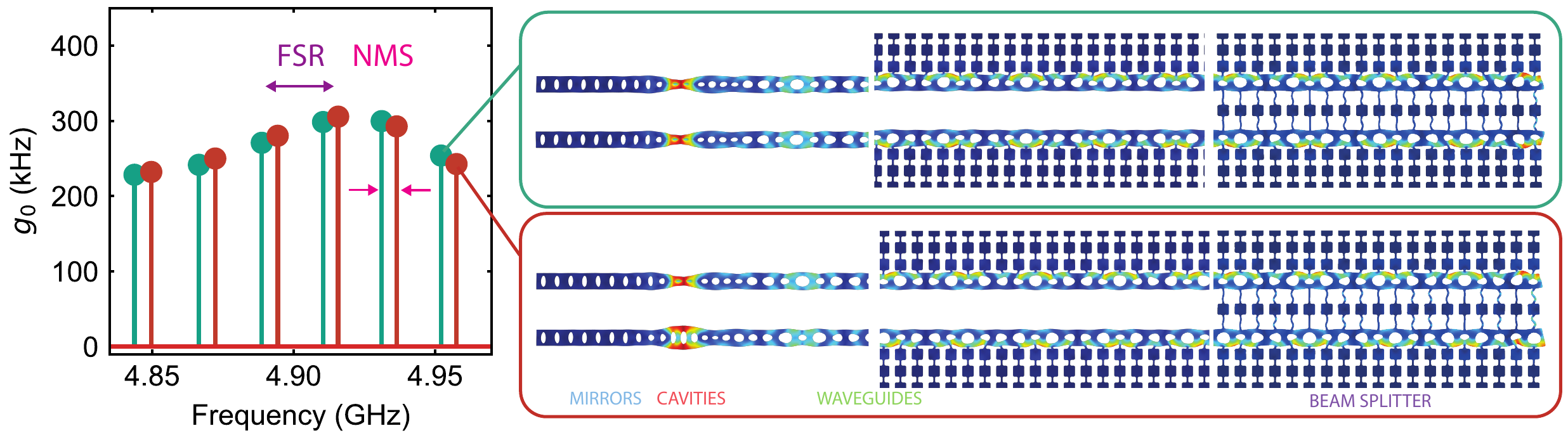}
	\caption{Simulation of the whole structure for a device with 50 unit cell of directional coupler and about \SI{80}{\micro m} total length (shorter than the measured device in the main text, for a faster simulation time). Left:\ calculated single photon optomechanical coupling rate for symmetric (green) and anti-symmetric (red) modes - for each Fabry-P\'erot mode, the symmetric and anti-symmetric modes form a doublet split by $\NMS$ due to the phononic coupling between the top and bottom waveguides. Right:\ mechanical mode profiles corresponding to one of the doublets. The whole structure is reported in segments for space reason. }
	\label{Fig:sim}
\end{figure*}

In Fig.~\ref{Fig:sim}, we show the mechanical simulation of the whole structure and the calculated optomechanical single photon coupling rate $g_{0}$ for different symmetric and anti-symmetric supermodes. The mechanical spectrum is a series of Fabry-P\'erot modes, equally spaced by $\FSR$, given by the hybridization of the single mode of the cavity with the series of modes supported by the free-ending waveguide. Each Fabry-P\'erot mode is split by $\NMS$ into a doublet due to the mechanical coupling between the two waveguides. The mode profiles corresponding to one of the doublets are shown in green and red boxes in Fig.~\ref{Fig:sim}. Note that these simulations use a shorter device length compared to the measured device, for faster computational run-times, which results in different $\NMS$ and $\FSR$. 

We use the full-width at half maximum of the distributions of $\NMS$ and $\FSR$ between the modes from the simulations as a metric to optimize the structure. We observe that changing the hole dimensions from the uncoupled waveguide region to the coupler region gives broader distributions of $\NMS$ and $\FSR$ that distorts the time-domain behavior of the wavepacket.

\section{Fabrication}
\label{SI:fab}
	
The device is fabricated from a silicon-on-insulator (SOI) wafer with a \SI{250}{nm} thick silicon device layer on top of \SI{3}{\mu m} of buried oxide. We use electron beam lithography to pattern the structure and transfer the mask with a dry HBr/Ar plasma etch. After processing, the chip is diced in order to access the device's optical waveguide with a lensed fiber in the dilution refrigerator. We then remove any remaining resist using \SI{80}{\degree} Dimethylformamide. Finally, the device is cleaned with a piranha solution and released using a 40\% hydrofluoric acid (HF) wet etch to remove the buried oxide layer~\cite{Riedinger2016}. In order to minimize oxidation, the device is immediately transferred into the dilution refrigerator.

Since we use a wet etch process to undercut the samples, we need to have very robust nanostructure. To achieve this, we use a thin ($<$50~nm wide) tether to connect the central optical waveguide (shared between the two cavities) and the optomechanical resonators. We report that using two sets of clamps on the optical waveguide (spaced by $\sim\SI{10}{\micro m}$) without the tether gives a yield of $\sim$70\%. We report no significant changes in the mechanical spectrum with or without the thin tether.

\section{Lifetime}
\label{SI:lifetime}

\begin{figure}[h]
	\includegraphics[width = 1 \linewidth]{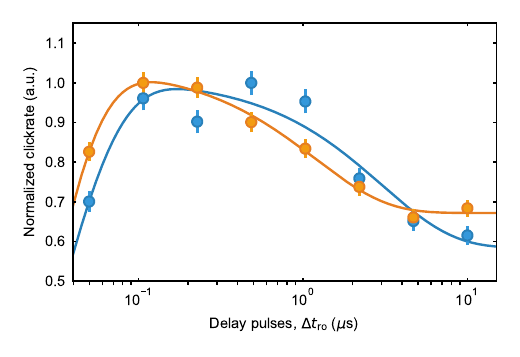}
	\caption{Normalized clickrates measured by the second probe pulse given by the thermally excited mechanical mode as a function of delay time $\Delta t_{ro}$ between pulses, for device A (blue) and B (orange). The population exhibits a double exponential behavior, given by the delayed absorption heating and the population decay. The measured lifetimes of the two cavities are $T_{1}\approx\SI{3.0}{\mu s}$ and $\SI{1.2}{\mu s}$ for cavity A and B, respectively.}
	\label{Fig:therm}
\end{figure}

In order to measure the mechanical lifetime $T_{1}$ of our devices, we send a sequence of two red-detuned pulses with varying delays $\Delta t_{ro}$. The first one is a strong pulse, used to heat our device through residual optical absorption, creating a large thermal population in the mechanical mode. The second pulse is weaker, allowing us to read the mechanical population after a variable delay from the first pulse. The results are shown in Fig.~\ref{Fig:therm} for both devices (A in blue and B in orange). As previously studied, the curves follow a double exponential decay~\cite{Wallucks2020}. We measure values of $T_{1} \approx \SI{3.0}{\mu s}$ and $\SI{1.2}{\mu s}$ for cavity A and B, respectively. These values are much longer than the typical round-trip time of the phononic wavepackets. They can be increased to several ms by adding additional phononic shield periods both at the waveguides' support part and the left end of the mirror side (blue part in Fig.~\ref{Fig:1}e), and values up to \SI{5.5}{ms} have been reported previously~\cite{Chan2012,Zivari2022,zivari2022chip}. We intentionally design and choose a device with short mechanical lifetime to increase the repetition rate of the experiment. The rise in the thermal population for short delays ($\Delta t_{ro} < \SI{100}{ns}$) is given by the delayed absorption~\cite{Riedinger2016}. 

The measured phonon lifetime directly reflects the phonon losses in our device. We can estimate a phonon loss coefficient $\gamma$ as $1/(v_{\rm g}T_1)$, where $v_{\rm g}$ is the phonon group velocity. We find $\gamma=83 (200)$~Np/m for device A (B), using $v_g=4,000$~m/s as extracted from the band structure simulations. This value, however, includes both loss mechanisms given by the total propagation loss inside the waveguide and beam splitter, and the optomechanical cavity. Extracting the contribution of the beam-splitter region only would require measuring devices with different length, and increased number of phononic shields at the left end of optomechanical cavity in Fig.~\ref{Fig:1}e (similar to~\cite{Zivari2022}), and goes beyond the scope of this work. The reported value sets only an upper limit.

\section{Thermal occupation of mechanical state}
\label{SI:nth}

\begin{figure}[h]
	\includegraphics[width = 1 \linewidth]{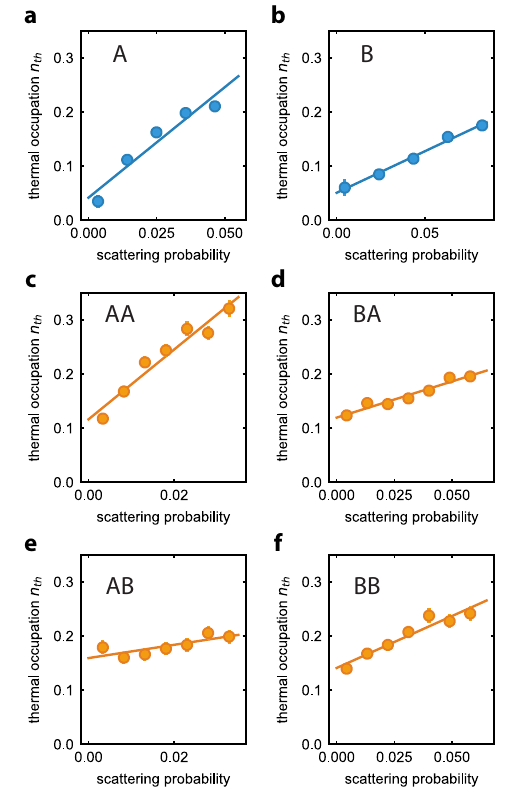}
	\caption{a,b) Thermal occupation of device A and B using a single pulse scheme versus the scattering probability of the pulse. c-f) Thermal occupation of device A and B in a two pulse sequence scheme, mimicking the realistic case of the experiment, versus the scattering probability of the pre-pulse. The pre-pulse is sent to one of the cavities, and the thermal occupation of either the same cavity or the opposite cavity is measured. Different combinations are specified with the two letters at the top left corner of each plot, with the first letter indicating the cavity that the pre-pulse is sent to, and the second letter the cavity that is measured.}
	\label{Fig:nth}
\end{figure}

The GHz mechanical excitations have in principle a thermal occupation of $<10^{-5}$ at mK temperatures. However, due to optical absorption of the pump laser in the device, a thermal mechanical population inside the cavity is created, and thus the thermal occupation of the mechanical mode is typically significantly higher. We measure the equivalent thermal occupation of the mechanical mode using the sideband thermometry technique~\cite{Riedinger2016,zivari2022chip,Zivari2022}. In particular, we send alternating blue-detuned and red-detuned pulses with equal energies (equal scattering probability), and use the click rates measured from both pulses to obtain the thermal occupation at the mode of interest~\cite{Riedinger2016}. We repeat this measurement with different pulse energies (varying the scattering probability). In this way we measure the amount of heating and thermal occupation induced by one single pulse in different cavities. The results are shown in Fig.~\ref{Fig:nth}a for device A and Fig.~\ref{Fig:nth}b for device B. Furthermore, we measure the thermal occupation in a two pulse sequence scheme, as it is used in our final experiment (see Fig.~\ref{Fig:4}), using the same technique with an additional red-detuned pre-pulse, \SI{115}{ns} before the pulses used to measure the thermal occupation. This red-detuned pre-pulse is used to mimic the heating of the blue pulse in the experiment. In this measurement we keep the energy of the sideband asymmetry pulses constant and equal to the value that we use in the actual experiment (\SI{280}{fJ}, scattering probabilities \SI{1.2}{\%} (\SI{2}{\%}) for cavity A (B)), and only vary the energy of the pre-pulse. We repeat this measurement in different combinations of sending the pre-pulse to a cavity (A or B) and measure the thermal occupation of one of the cavities (A or B). The results for different combinations are shown in Fig.~\ref{Fig:nth}c-f. These measurements show that by sending a pulse to one cavity, not only do we heat that cavity, but some of the thermal energy also heats the other cavity through the coupler. This is clearly visible from the linear increase of the thermal population with the scattering probability of the pre-pulse and by the higher thermal population compared to the single pulse case (as in Fig.~\ref{Fig:nth}a and b). Note that, since the cavities have optical resonances spaced by much more than the optical linewidth and the mechanical frequencies of interested, the optical pulses that address one cavity does not interact at all with the other, causing no direct heating. For the pulse energies used in the actual experiment, we measure $n_{th,AA} \approx 0.18$, $n_{th,AB} \approx 0.17$, $n_{th,BA} \approx 0.14$ and $n_{th,BB} \approx 0.17$.

\begin{figure*}[ht!]
	\includegraphics[width = 1.0 \linewidth]{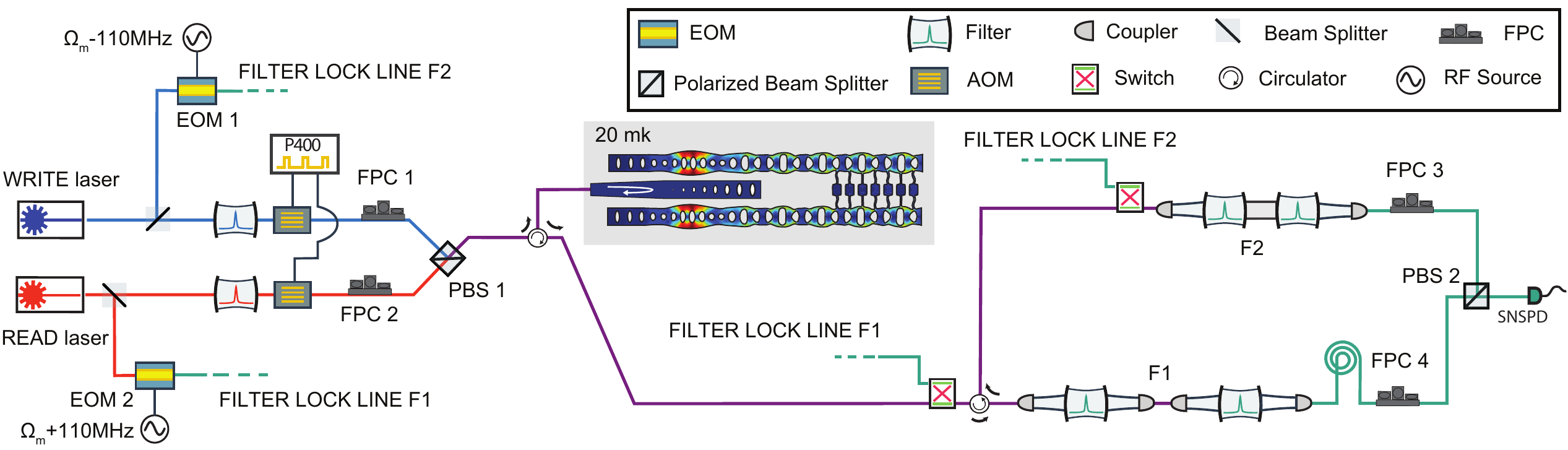}
	\caption{Detailed sketch of the experimental setup. P400 is the digital pulse generator to create pulses on the acousto-optic modulators (AOMs). EOMs are electro-optic modulators, BS the beam splitter, PBS the polarized beam splitter, FPC the fiber polarization controller and SNSPD the superconducting nanowire single photon detector. $\Omega_{m}$ is the mechanical central frequency (\SI{5.31}{GHz}). We have two sets of free-space detection filters F1 and F2, in order to distinguish the light from different devices A and B, when we excite and read different cavities (see the text for more details).}
	\label{Fig:setup}
\end{figure*}

\section{Setup}
\label{SI:setup}

A sketch of the experimental setup is depicted in Fig.~\ref{Fig:setup}. We use two continuous-wave lasers for write and read operations, whose frequencies are locked using active feedback from a wavelength meter. We then filter the GHz noise of the lasers using fiber filters. Both lasers are pulsed by gating \SI{110}{MHz} acousto-optic modulators (AOMs) with a digital pulse generator (P400). The two lines are combined on a polarized beam splitter (PBS1), and we use fiber polarization controllers (FPCs) on each line to maximize transmission through PBS1. After recombination, the pulses are sent into a lensed fiber at the mixing chamber of the dilution refrigerator, coupling the optical fields into the device. The coupling efficiency from the fiber to the central optical waveguide (yellow part in Fig.~\ref{Fig:1}e) is $\eta_{\rm c} \approx \SI{40}{\%}$. The reflected light from the device is routed to free-space filter cavities (F1 and F2) using a circulator, in order to filter the write and read pump pulses with suppression ratio of \SI{98}{dB} for F1 and \SI{96}{dB} for F2. The efficiency of the filter setup for the Stokes and anti-Stokes scattered photons is $\eta_{f} \approx \SI{30}{\%}$. This suppression corresponds to a pump tone leakage of approx.\ $2-5 \times 10^{-6}$ clicks per trial over the total Stokes and anti-Stokes pulse area. This is at least one order of magnitude lower than the Stokes and anti-Stokes click rates of $2 \times 10^{-4}$ and $4 \times 10^{-5}$, respectively. When we write on one device and read the same (AA or BB combinations), using one filter cavity (F1) suffices for our detection. However, when we write on one device and read the opposite device (AB or BA combinations), we need to use both filter setups (F1 and F2), since the devices have resonances at different wavelengths. In this scheme, F1 is locked on resonance of one device (A or B) while F2 is locked to the other resonance (B or A). The anti-Stokes optomechanically scattered photon from the read pulse passes through F1, while all the pumps and the Stokes photon are reflected back into fiber. The light is routed to F2 using a circulator which suppresses all pump fields and only allows the Stokes photon to pass. Every \SI{8}{s} the measurement is paused and locking lasers are sent to the cavities to lock them simultaneously. We use two EOMs to create sidebands at the resonances wavelengths using light from the read laser and from the write laser to lock F1 and F2, respectively. We merge both signals from F1 and F2 on a polarizing beam splitter (PBS), and use FPCs to maximize transmission of each signal through the PBS. We send the output of PBS to a SNSPD installed on the quasi \SI{1}{K} stage of our dilution refrigerator. Since the filters are being locked to two different wavelengths, we observe relatively high leakage from the pump lasers of opposite wavelength -- the write laser pump leaks through F1 and the read laser pump leaks through F2. This is because the suppression of the filter setup depends heavily on the detuning between the filter resonances and the laser wavelength. Given the (fixed) free spectral range of the filter cavities, the opposite laser can have a smaller detuning and so smaller suppression compared to the laser used to lock the cavity. In order to temporally distinguish between the leakage of the pump lasers and the actual Stokes and anti-Stokes scattered photons coming out of the filter setup outputs, we delay the signal of F1 (the one we use for reading) by \SI{320}{ns}. In this way, the leakages are spaced in time from the Stokes and anti-Stokes photons more than the recovery time of the SNSPDs (which is about \SI{100}{ns}), and they do not affect the measurement results. Additionally, we measure a dark count rate of about $10^{-6}$ clicks per trial, integrated over the \SI{30}{ns} pulse area.

\section{Coherent drive measurement}	
\label{SI:coh_drive}

For the coherent drive measurement, we first gather the click rates given by our quasi-continuous red-detuned read laser. Because of the optical absorption of the device, the reading laser creates some thermal occupation in the cavity as well. To independently calibrate this background and measure only the coherent population of our cavities, we run a measurement with the same reading power on both cavities, without the coherent excitation pulse. This gives us the thermal background induced by the reading laser, which we then subtract from the original data. In order to calculate the normalized phononic population, we first normalize the click rates by the maximum click rates we get from the excitation pulse (the peak at zero time delay). We also normalize the click rates with the optomechanical coupling rate $g_{0}$, as well as the extrinsic optical couplings $\kappa_{e} = \kappa - \kappa_{i}$ of each cavity (see details in Sec.\,\ref{sec:coh_drive_theory}). In this way, we normalize the phononic population with the phononic excitation and read-out efficiency.

\begin{figure}[h!]
	\includegraphics[width = 1.0 \linewidth]{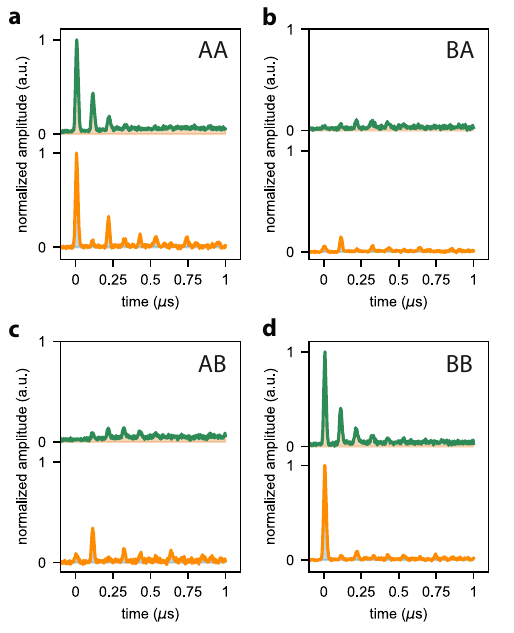}
	\caption{Normalized coherent phononic population of each cavity when either the same cavity (a and d) or opposite cavities (b and c) are excited. The green (orange) plot on top (bottom) corresponds to a device consisting of 25 (100) unit cells in the phononic coupler region.}
	\label{Fig:coh_drive_SI}
\end{figure}

The measurements for devices with 25 and 100 unit cells in the beam splitter are shown in Fig.~\ref{Fig:coh_drive_SI} for different exciting and reading combinations. As can be clearly seen, the 25 unit cell device (green) shows behavior of a \SI{80}{\%} transmissive beam splitter, whereas the 100 unit cell device (orange) of a beam splitter with almost \SI{100}{\%} reflectivity. This behavior follows our simulations very closely. The decay in the amplitude of the peaks is mainly given by the imperfect rephasing of the population given by the imperfect mechanical spectrum. Imperfections in the mechanical spectrum are caused by residual dispersion of the waveguides and beam splitters, coming from fabrication disorder. The phononic dissipation is negligible for the delays analyzed in the paper~\cite{Zivari2022}.

We can use the coherent drive measurements to estimate potential reflections occurring at the interface between the uncoupled waveguides and the coupler region due to the abrupt change in the phononic crystal structure. Assuming reflections at the coupler interface, we would end up with a cavity of length $L = L_{\rm wg} – L_{\rm c}$ , with $L_{\rm wg}$ the full length of the device and $L_{\rm c}$ the length of the coupler. For the 50:50 splitter case, shown in Fig.\,3b of the main text, we get $L$ = \SI{183.7}{um}, and given a group velocity of \SI{4000}{m/s}, we expect a first bouncing peak about \SI{92}{ns} after the excitation pulse, while the directional coupler signal should give the first bouncing at \SI{105}{ns}. The pulse from direct reflection is not clearly visible from the data. Given the pulse length of \SI{30}{ns}, we can be even more precise by looking at the data in Fig.\,\ref{Fig:coh_drive_SI}, which shows the bouncing pattern for the longest coupling region we measured. This would give $L$ = \SI{157.4}{um}, resulting in a bouncing peak \SI{79}{ns} after the excitation pulse, which would almost be completely separated from the coupler first bouncing. There is not experimental evidence of this behavior, and we conclude that direct reflections are negligible.

\subsection{Transfer matrix approach}
To model the data shown in Fig. 3c, we use a transfer matrix approach. The transmission through the directional coupler is modeled via a lossless beam splitter matrix $\Ubs$:
\begin{equation}
	\Ubs = \begin{pmatrix}
		\sqrt{T} & i\sqrt{R} \\ 
		i\sqrt{R}&  \sqrt{T} 
	\end{pmatrix}
	\label{eq:ubs}
\end{equation}
with $R=|r|^2$ and $T=|t|^2$, and $r, t$ defined as in the main text. Given the round-trip time $\taurt$, and defining $A(t)$ and $B(t)$ as the phononic population amplitude in device A and B as a function of time, respectively, we can write:
\begin{equation}
	S(( n+1 ) \taurt ) = \Ubs S(n \taurt)
\end{equation}
with $n$ and integer number, $S(t) = (A(t) ,B(t))^T$, and $T$ the transpose operator. After $N$ round-trips, the phononic population in the devices can be simply calculated as $S(N \taurt) = (\Ubs)^N S(0)$. After we obtain the amplitudes for each round trip, we multiply all round-trip populations by a common factor $A_{ij}$, with $i,j = {\rm A,B}$ independently for each combination of excitation/detection. As explained in the main text, this factor takes into account losses and dispersion mismatch between devices. The only condition we impose is $A_{\rm AB} = A_{\rm BA}$ to satisfy the reciprocity condition.

\subsection{Theory of coherent drive \label{sec:coh_drive_theory}}
Here we present the theory of the coherent drive measurement in our system that we use to normalize and extract the mechanical population in our devices from the optical click rates. We divide the optical fields into two different fields, one corresponding to the drive field and one corresponding to the readout field - the anti-Stokes scattered photons used to map the mechanical state onto a photonic state. Let's first consider the drive field. We start from the optomechanical Hamiltonian in its general form in the rotating frame of the laser field:
\begin{equation}
	H = \hbar \Delta a^{\dagger}a + \hbar \Omega b^{\dagger}b + \hbar g (a^{\dagger} + a)(b^{\dagger} + b) 
\end{equation}
with $a$ and $b$ representing annihilation operators for optical and mechanical field, $\Delta = \omega_{c} - \omega_{L}$ the laser detuning from the optical cavity frequency, $\Omega$ the central mechanical frequency, and $g$ the equivalent collective optomechanical coupling of the wavepacket. We consider input laser field with the EOM driven sidebands as $a_{in}(1 + 2\alpha cos(\Omega t))$, and thus the Langevin equations of the motion will follow:
\begin{align}
	&\frac{da}{dt} = -(\kappa / 2 + i\Delta)a + \sqrt{\kappa_{e}}a_{in}(1+2\alpha cos(\Omega t))
	\label{dadt}
	\\
	&\frac{db}{dt} = -(\Gamma /2 + i\Omega)b - ig(a + a^{\dagger})
	\label{dbdt}
\end{align}
in which $\kappa$ and $\Gamma$ represent the optical cavity and mechanical loss, and $\kappa_{e}$ the extrinsic coupling of the optical cavity to the optical waveguide. Here we neglect the optomechanical coupling in Eq.~\eqref{dadt} due to the large coherent laser field. Moreover is Eq.~\eqref{dbdt} we neglect any input mechanical field such as thermal occupation and assume that both fields are dominated by the coherent populations. Assuming the steady state where the fields oscillate with the drive frequency $\Omega$, we assume fields as $j = j_0 + j_- e^{-i\Omega t} + j_+ e^{i\Omega t}$ where $j = {a,b}$. By solving the substituting these values in Eq.~\eqref{dadt} and \eqref{dbdt} and applying the approximation $\Omega \gg \Gamma$, we obtain the mechanical field as:
\begin{equation}
\begin{split}
\widetilde{b(t)} = &\frac{-ig}{\Gamma / 2}.[\frac{1}{\kappa / 2 + i(\Delta - \Omega)} + \frac{1}{\kappa / 2 - i(\Delta + \Omega)}]. \\
&\alpha\sqrt{\kappa_e}a_{in}. e^{-i\Omega t}
\end{split}
\end{equation}
In order to read the mechanical state, we send a continuous laser red-detuned by the mechanical frequency from the optical cavity to continuously read the mechanical population inside the cavity. In this way we get the state-swap interaction Hamiltonian of below in the rotating frame of the laser field:
\begin{equation}
H = \hbar \Omega a^\dagger a + \hbar \Omega b^\dagger b + \hbar g (a^\dagger b + ab^\dagger)
\end{equation}
In this regime, the Langevin equations of motion would follow:
\begin{align}
&\frac{da}{dt} = -(\kappa / 2 + i\Omega)a - igb 
\label{dadt_read}
\\
&\frac{db}{dt} = -(\Gamma / 2 + i\Omega)b + b_{in}
\label{dbdt_read}
\end{align}
Here we only considered the first order perturbation term of the optical field, therefore neglecting the laser field and only considering the Anti-Stokes scattered photons on-resonance with the optical cavity at frequency $\omega_c$. Moreover, we use the mechanical field in Eq.~\eqref{dbdt_read} as an input drive field in Eq.~\eqref{dbdt_read}. We assume this field as $\sqrt{\eta} \widetilde{b(t)}$ in which $\eta$ mimics the energy flow of the phononic wavepacket in different round trips, including the splitting, dispersion and dissipation effects altogether. We thus can assume the optical field oscillating similarly as the mechanical frequency $a = a_-^{'} e^{-i\Omega t}$ and using the input-output theory, we can extract the output field of Anti-Stokes scattered photons as:
\begin{equation}
a_{out} = \frac{ig\sqrt{\eta} \sqrt{\kappa_e}}{\kappa / 2 . \Gamma / 2}.\widetilde{b(t)}
\end{equation}
Ultimately, by detecting the optical field on the SNSPDs we measure $<a_{out}^\dagger a_{out}>$.

For normalization of the field, we have to take into account that different devices on each pair, have different $\kappa_e$ and $g$. For the case where we excite and measure on the same device, we can easily extract $\eta$ by normalizing the click rates to the one of the $0^{th}$ peak from the excitation pulse. However, for the case where we excite and read different devices, after normalizing to the $0^{th}$ peak of the excitation, we also need to consider the additional factor coming from different $g$ and $\kappa_e$ of the devices. Therefore, for instance if we excite device A and read device B, we need to multiply the result by $\frac{g_A^2 . \kappa_{e,A}}{g_B^2 . \kappa_{e,B}}$.

\section{Two photon cross correlation $g^{(2)}_{om}$ versus time}
\label{SI:gcc}

\begin{figure}[t!]
	\includegraphics[width = 1.0 \linewidth]{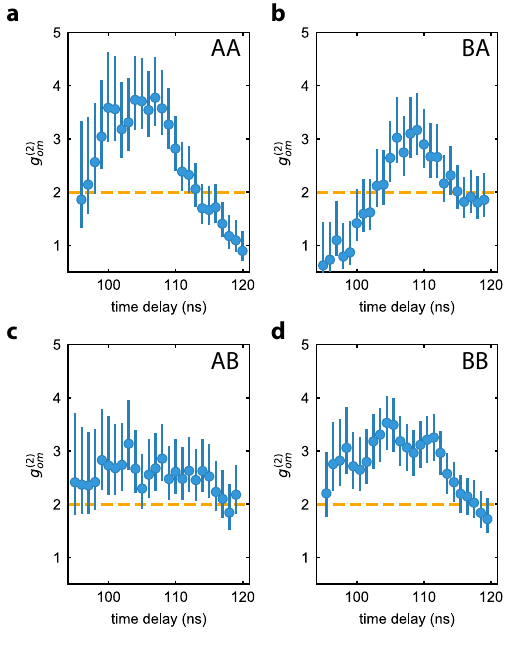}
	\caption{The two-photon cross correlation $g^{(2)}_{om}$ behavior in time for the 50:50 beam splitter device. The orange dashed line indicates the classical boundary of $2$.}
	\label{Fig:gcc_SI}
\end{figure}

In order to measure the two photon cross correlation of the phononic wavepackets, we use a technique similar to~\cite{Zivari2022} to filter out coincidences in time with finer resolution. This allows us to obtain the time domain behavior of the phononic packet. The results for different combinations of the excite and read cavities are shown in Fig.~\ref{Fig:gcc_SI}. The correlation value increases from uncorrelated (1) or classical correlation (2) to non-classical ($>$2) value as the phononic wavepacket returns back to the optomechanical cavity, and again drops down to classical correlation (2) and eventually uncorrelated values as the packet leave the cavity to the phononic waveguide again.

\end{document}